\documentclass[aps,twocolumn,showpacs]{revtex4}
\usepackage{epsfig}
\usepackage{amsbsy}
\usepackage{amsmath}
\usepackage{amsfonts}
\usepackage{graphicx}
\usepackage{rotating}
\usepackage{color}

\begin{document}

\newcommand{\nn}{\nonumber \\}
\newcommand{\bra}[1]{\langle{#1}|}
\newcommand{\ket}[1]{|{#1}\rangle}
\newcommand{\braket}[2]{\langle{#1}|{#2}\rangle}
\newcommand{\dg}{^\dagger}
\newcommand{\h}{\mathcal{H}}
\newcommand{\e}{\mathcal{E}}
\newcommand{\Tr}{\textrm{Tr}}
\newcommand{\anc}{\textrm{anc}}
\newcommand{\s}{s}

\title{Lower bounds for communication capacities of two-qudit unitary operations}

\author{Dominic W.\ Berry}\thanks{
Electronic address: {\tt dberry\textcolor{white}{a}\!\!\!%
{@}\textcolor{white}{a}\!\!\!%
ics.mq.edu.au}}
\affiliation{Centre for Quantum Computer Technology, Macquarie University,
Sydney, NSW 2109, Australia}
\affiliation{Department of Physics, The University of Queensland, St.\ Lucia,
	Queensland 4072, Australia }

\begin{abstract}
We show that entangling capacities based on the Jamio{\l}kowski isomorphism may
be used to place lower bounds on the communication capacities of arbitrary
bipartite unitaries. Therefore, for these definitions, the relations which
have been previously shown for two-qubit unitaries also hold for arbitrary
dimensions. These results are closely related to the theory of the
entanglement-assisted capacity of channels. We also present more general methods
for producing ensembles for communication from initial states for entanglement
creation.
\end{abstract}
\date{\today}

\pacs{03.67.Mn, 03.67.Hk}

\maketitle

\section{Introduction}
In quantum information processing, one of the most important tasks is to perform
operations between different subsystems. It is therefore important to quantify
what resources are required for implementing these operations
\cite{cirac,dur,implement,eisert,collins}, and conversely the resources that
these operations are capable of creating
\cite{zanardi,durvid,kradur,kraus,leifer,childs,kraus2,wang,bravyi,groisman}.
In the very simple cases of SWAP and CNOT operations, many of
the capacities are equal, or related by integer ratios \cite{eisert,collins}. 
For more general bipartite unitary operations, a number of relations may be
shown \cite{bhls}, and further relations may be shown for two-qubit unitaries
\cite{berry1,berry2,berry3}. In the case of two-qudit (multilevel) systems, it
has been shown that many symmetries and inequalities do not hold
\cite{linden,aram}.

One relation that holds for all bipartite unitaries is that if they create
entanglement then they can also perform classical communication \cite{bhls}.
The method used in Ref.\ \cite{bhls} is somewhat indirect, and only gives a
small lower bound on the communication that can be achieved. Here we give a
more useful lower bound, which demonstrates that a significant amount of
communication may be achieved.

In the case of two-qubit unitaries, the communication that can be achieved in
each direction is at least as large as the entanglement that can be created
\cite{berry1,berry2}. The derivation used in Refs.\ \cite{berry1,berry2} does
not apply to the more general case of two-qudit unitaries, and a counterexample
for the two-qudit case is given in Ref.\ \cite{aram}.

In Ref.\ \cite{berry3} it was shown that, in the restricted case of controlled
unitary operations, the entanglement capacity again gives a lower bound on the
communication capacity. Here we show here that it is possible to derive a
similar inequality for general two-qudit unitaries provided we use an
alternative definition of the entangling capacity.

This paper proceeds as follows. The definitions for the entangling and Holevo
capacities are presented in Sec.\ \ref{sec:def}. In Sec.\ \ref{sec:rel} the
lower bounds on the Holevo capacities in terms of the entangling capacities are
shown. The relation between these results and those for quantum channels is
discussed in Sec.\ \ref{sec:cha}. In Sec.\ \ref{sec:imp} we give further
generali{\s}ations of these results, and discuss the consequences of the results
in Ref.\ \cite{aram}. Conclusions are given in Sec.\ \ref{sec:con}.

\section{Definitions}
\label{sec:def}
We subdivide the system into two subsystems, $\h_A$ and $\h_B$, that
are the subsystems in the possession of Alice and Bob, respectively. We further
subdivide these subsystems as
\begin{equation}
\label{ancilla}
\h_A = \h_{A_\anc} \otimes \h_{A_U}, \quad
\h_B = \h_{B_U} \otimes \h_{B_\anc}.
\end{equation}
The subsystems labelled ``anc'' are the ancillas, and those labelled ``$U$'' are
the subsystems that the unitary operation $U$ acts upon.

\subsection{Entangling capacities}
\label{sec:ent}
There are two main definitions of the entangling capacity of a unitary operation
that have been presented in preceding work \cite{bhls,leifer}:
\begin{align}
\label{eq:eu}
E_U &\equiv \sup_{\ket{\phi}_A\in\h_A,\ket{\chi}_B\in\h_B}
E(U\ket{\phi}_A\ket{\chi}_B) \\
\label{eq:chan}
\Delta E_U &\equiv \sup_{\ket{\psi}_{AB}\in\h_A\otimes\h_B}
\left[ E(U\ket{\psi}_{AB}) - E(\ket{\psi}_{AB})\right].
\end{align}
The quantity $E(\cdots)$ is the entropy of entanglement
$E(\ket{\psi}) = S[\Tr_A(\ket{\psi}\bra{\psi})]$, where
$S(\rho)=-\Tr(\rho \log\rho)$. Throughout we employ logarithms to base 2,
so the entanglement is expressed in units of ebits. The capacity $E_U$
corresponds to the maximum entanglement that may be achieved starting with
states that are not entangled between $\h_A$ and $\h_B$, but may be entangled with
ancillas. The second capacity, $\Delta E_U$, is the maximal increase in
entanglement. The capacity $\Delta E_U$ may also be related to the average
entanglement which may be created asymptotically \cite{bhls}.

We may add to these definitions
\begin{align}
E_U^{\Psi} &\equiv E(U\ket{\Psi}_A\ket{\Psi}_B), \\
\label{eq:difen}
\Delta E_U^{\Psi} &\equiv \max_{U_0} [E_{UU_0}^{\Psi} - E_{U_0}^{\Psi}].
\end{align}
where $\ket{\Psi}$ is the maximally entangled state $\sum_{j}\ket{jj}$ and $U_0$
is a unitary operation on $\h_{A_U}\otimes\h_{B_U}$. The definition of $\Delta
E_U^{\Psi}$ is equivalent to that of $\Delta E_U$, except we have restricted to
states $\ket{\psi}_{AB}$ of the form $U_0\ket{\Psi}_A\ket{\Psi}_B$. We may take
the maximum rather than the supremum, because this is a continuous function on a
compact set. Clearly these two definitions are more restrictive than
Eqs.\ \eqref{eq:eu} and \eqref{eq:chan}, so we have $E_U^{\Psi}\le E_U$ and
$\Delta E_U^{\Psi} \le \Delta E_U$. Also, because we may take $U_0$ to be the
identity, $E_U^{\Psi}\le \Delta E_U^{\Psi}$.

As motivation for these definitions, for a range of two-qubit unitaries
the equalities $E_U^{\Psi}=E_U$ and $\Delta E_U^{\Psi}=\Delta E_U$ hold
\cite{leifer,berry3}. In addition, $U\ket{\Psi}_A\ket{\Psi}_B$ is the state
that is equivalent to $U$ under the Jamio{\l}kowski isomorphism \cite{jam}.
That is, $U\ket{\Psi}_A\ket{\Psi}_B$ may be used to implement $U$ by
postselecting on particular Bell measurement results. The probability of
obtaining the correct measurement results is $(d_{A_U}d_{B_U})^2$.

\begin{figure}
\centering
\includegraphics[width=0.45\textwidth]{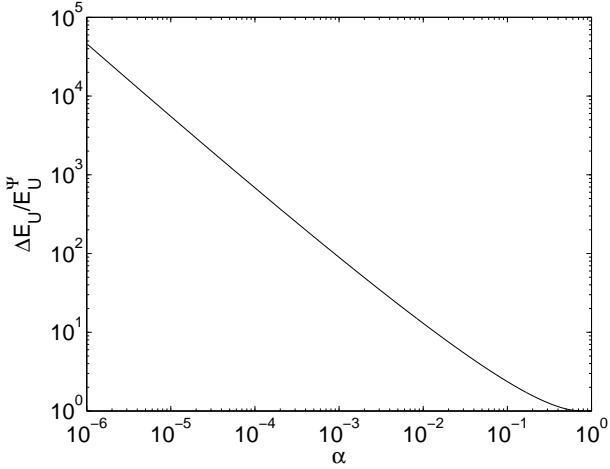}
\caption{The ratio of $\Delta E_U$ to $E_U^{\Psi}$ for the operation $U(\alpha)
=\exp(i\alpha\sigma_z\otimes\sigma_z)$.}
\label{unbou}
\end{figure}

It is possible for the entanglement to increase for one particular measurement
result, so it is possible for $E_U^{\Psi}$ to be strictly less than $E_U$.
However, the entanglement when averaged over all possible measurement results
can not increase, so we must have
\begin{equation}
E_U^{\Psi}(d_{A_U}d_{B_U})^2 \ge E_U.
\end{equation}
We do not obtain a similar relationship between $E_U^{\Psi}$ and $\Delta E_U$,
because if the initial target state has entanglement, the total entanglement can
decrease by more than $E_U^{\Psi}$ for some measurement results. In fact,
the ratio of $\Delta E_U$ to $E_U^{\Psi}$ is unbounded, as is easily seen for
the case of the unitary $U(\alpha)=\exp(i\alpha\sigma_z\otimes\sigma_z)$ (see
Fig.\ \ref{unbou}).

We also define entangling capacities with the state only
restricted on one side or the other:
\begin{align}
\label{uni1}
E_U^{\Psi,\rightarrow} &\equiv \sup_{\ket{\chi}_B} E(U\ket{\Psi}_A\ket{\chi}_B), \\
\label{uni2}
\Delta E_U^{\Psi,\rightarrow} &\equiv \sup_{\ket{\chi}_B,U_0}
[E(UU_0\ket{\Psi}_A\ket{\chi}_B)-E(U_0\ket{\Psi}_A\ket{\chi}_B)], \\
E_U^{\Psi,\leftarrow} &\equiv \sup_{\ket{\phi}_B} E(U\ket{\phi}_A\ket{\Psi}_B), \\
\label{uni4}
\Delta E_U^{\Psi,\leftarrow} &\equiv \sup_{\ket{\phi}_B,U_0}
[E(UU_0\ket{\phi}_A\ket{\Psi}_B)-E(U_0\ket{\phi}_A\ket{\Psi}_B)].
\end{align}
From the definitions it is clear that $E_U^{\Psi} \le E_U^{\Psi,*}\le E_U$
and $\Delta E_U^{\Psi}\le \Delta E_U^{\Psi,*}\le \Delta E_U$. Throughout this
paper we use the asterisk to indicate either $\leftarrow$ or $\rightarrow$.

Next we consider what happens if we allow the dimensions of $\h_{A_U}$ and
$\h_{B_U}$ to be expanded. That is, we may take $\h_{A_U}=\h_{A_U^2}
\otimes\h_{A_U^1}$ and $\h_{B_U}=\h_{B_U^1}\otimes\h_{B_U^2}$, and allow
$U$ to act upon $\h_{A_U}\otimes\h_{B_U}$, although it only acts as the identity
upon $\h_{A_U^2}$ and $\h_{B_U^2}$. In this case, we find that
$\Delta E_U^{\Psi,*}=\Delta E_U$.

For the example of $\Delta E_U^{\Psi,\rightarrow}$, the result may be shown in
the following way. We take the initial state
\begin{equation}
\ket{\Psi}_A \ket{\psi}_{B_U},
\end{equation}
where $\ket\psi$ is a state giving a change in entanglement within $\epsilon$ of
the supremum in \eqref{eq:chan}. We take
\begin{align}
\h_{A_U}&=\h_{A_U^2}\otimes\h_{A_U^1}, \nn
\h_{B_U}&=\h_{B_U^1}\otimes\h_{B_U^2}\otimes\h_{B_U^3}.
\end{align}
The dimension of $\h_{A_U^2}$ is taken to be equal to that of Alice's ancilla
in $\ket\psi_{AB}$ [with the subsystems as in the definition \eqref{eq:chan}],
the dimension of $\h_{B_U^2}$ is equal to that of Bob's ancilla in
$\ket\psi_{AB}$, and the dimension of $\h_{B_U^3}$ is equal to that of Alice's
component of $\ket\psi_{AB}$. The dimensions of $\h_{A_U^1}$ and
$\h_{B_U^1}$ are the same as $\h_{A_U}$ and $\h_{B_U}$ for $\ket\psi_{AB}$
(that is, they are the subsystems that $U$ acts nontrivially upon).

The operation $U_0$ is taken to swap the states of
the subsystems $\h_{A_U}$ and $\h_{B_{U,3}}$, giving
\begin{equation}
\ket{\Psi}_{A_\anc,B_U^3} \ket{\psi}_{A_U,B_U^1,B_U^2}.
\end{equation}
Because $U$ only acts nontrivially on $\h_{A_U^1}\otimes\h_{B_U^1}$, the change
in entanglement under $U$ is $\Delta E_U-\epsilon$. As we may take $\epsilon$ to
be arbitrarily small, $\Delta E_U^{\Psi,\rightarrow}=\Delta E_U$.

All communication capacitites considered here are unaltered if larger dimensions
are permitted for $\h_{A_U}$ and $\h_{B_U}$. Therefore, any relations that can
be proven for $\Delta E_U^{\Psi,*}$ also hold for $\Delta E_U$. The only other
capacity which we consider in this work that may depend on the dimensions of
$\h_{A_U}$ and $\h_{B_U}$ is $\Delta E_U^\Psi$. However, in this case it does
not appear that it simplifies to $\Delta E_U$ under an expanded dimension.

\subsection{Holevo capacities}
Next we define a range of capacities for increasing the Holevo information. An
ensemble is a set of states $\{\rho_j\}$ that are supplied with
probabilities $p_j$. The ensemble is
denoted by $\e=\{ p_j,\rho_j \}$. 
The Holevo information of the ensemble $\e$ is given by \cite{kholevo}
\begin{equation}
\chi(\e) = S(\bar\rho) - \sum_j p_j S(\rho_j),
\end{equation}
where $\bar\rho=\sum_j p_j \rho_j$. This quantity gives the asymptotic
communication that may be performed using an ensemble by coding over
multiple copies \cite{holevo,shuwes}.

The Holevo capacities are defined by \cite{bhls,berry1,berry2,berry3}
\begin{align}
\label{hol1}
\chi_U^{\rightarrow} &\equiv \sup_{ p_j, V_j^A, \ket{\psi}_{AB} }
\chi\left( p_j, \Tr_A U V_j^A \ket{\psi}_{AB} \right), \\
\chi_U^{\leftarrow} &\equiv \sup_{ p_j, V_j^B, \ket{\psi}_{AB} }
\chi\left( p_j, \Tr_B U V_j^B \ket{\psi}_{AB} \right), \\
\Delta\chi_U^{\rightarrow} &\equiv \sup_\e \big[ \chi ( \Tr_A U\e )
- \chi ( \Tr_A \e ) \big], \\
\label{hol4}
\Delta\chi_U^{\leftarrow} &\equiv \sup_\e \big[ \chi ( \Tr_B U\e )
- \chi ( \Tr_B \e ) \big].
\end{align}
Here $V_j^A$ and $V_j^B$ are local unitaries that act in Alice and Bob's
subsystems, respectively. Throughout we use the convention that the
superscripts on local operators indicate the subsystems they act upon. We also
use the notation conventions 
\begin{align}
\Tr_X(\ket{\Phi}) &\equiv \Tr_X\ket{\Phi}\bra{\Phi}, \\
U\e &\equiv \{ p_j, U\ket{\Phi_j} \}, \\
\Tr_X\e &\equiv \{ p_j, \Tr_X(\ket{\Phi_j}) \}.
\end{align}
We also omit the curly brackets when directly taking the Holevo
information of an ensemble.

The capacities $\chi_U^{\rightarrow}$ and $\chi_U^{\leftarrow}$ are the
capacities for obtaining Holevo information on Bob and Alice's sides,
respectively, without initial correlations. The capacities
$\Delta\chi_U^{\rightarrow}$ and $\Delta\chi_U^{\leftarrow}$ are the
capacities for increasing the Holevo information on Bob and Alice's sides,
respectively. These capacities give the asymptotic communication capacities
from Alice to Bob and Bob to Alice, respectively \cite{bhls}. From the
definitions it is clear that $\chi_U^{*}\le \Delta\chi_U^{*}$
($*$ indicates $\rightarrow$ or $\leftarrow$).

One may also define Holevo capacities for initially unentangled states
\cite{berry3}:
\begin{align}
\chi_U'^{\rightarrow} &\equiv \sup_{ p_j, \ket{\phi_j}_{A}, \ket{\chi}_{B} }
\chi ( p_j, \Tr_A U \ket{\phi_j}_{A}\ket{\chi}_{B} ), \\
\chi_U'^{\leftarrow} &\equiv \sup_{ p_j, \ket{\phi}_{A}, \ket{\chi_j}_{B} }
\chi ( p_j, \Tr_B U \ket{\phi}_{A}\ket{\chi_j}_{B} ), \\
\Delta\chi_U'^{\rightarrow} &\equiv \sup_{ p_j, \ket{\phi_j}_{A}, \ket{\chi_j}_{B} }\big[
\chi ( p_j, \Tr_A U \ket{\phi_j}_{A}\ket{\chi_j}_{B} ) \\
& \qquad\qquad -\chi ( p_j, \Tr_A \ket{\phi_j}_{A}\ket{\chi_j}_{B} )\big], \\
\Delta\chi_U'^{\leftarrow} &\equiv \sup_{ p_j, \ket{\phi_j}_{A}, \ket{\chi_j}_{B} }\big[
\chi ( p_j, \Tr_B U \ket{\phi_j}_{A}\ket{\chi_j}_{B} ) \nn
& \qquad\qquad -\chi ( p_j, \Tr_B \ket{\phi_j}_{A}\ket{\chi_j}_{B} )\big].
\end{align}
These capacities are more restrictive than those in Eqs.\ \eqref{hol1} to
\eqref{hol4}, so $\chi_U'^{*}\le\chi_U^{*}$ and
$\Delta\chi_U'^{*}\le\Delta\chi_U^{*}$. It is also clear from the definitions
that $\chi_U'^{*}\le\Delta\chi_U'^{*}$.
In Ref.\ \cite{berry3} it was shown that $\chi_U'^{*}\le E_U$ and
$\Delta\chi_U'^{*}\le \Delta E_U$. In addition, it was shown that, for
controlled unitary operations, $\chi_U'^{*}=E_U$ and
$\Delta\chi_U'^{*}=\Delta E_U$.

\section{Relations between capacities}
\label{sec:rel}
\subsection{Symmetry}
The capacities $E_U^{\Psi}$ and $\Delta E_U^{\Psi}$ differ from the usual
capacities in that they are intrinsically time symmetric. That is,
\begin{equation}
E_U^{\Psi} = E_{U\dg}^{\Psi}, \qquad \Delta E_U^{\Psi}=\Delta E_{U\dg}^{\Psi}.
\end{equation}
To show this for $E_U^{\Psi}$, note that 
\begin{equation}
U\dg\ket{\Psi}_A\ket{\Psi}_B
= (U^{A_\anc B_\anc})^*\ket{\Psi}_A\ket{\Psi}_B.
\end{equation}
Here we use superscripts on $U$ to indicate that it is acting on the ancillas;
otherwise it is acting upon $\h_{A_U}\otimes \h_{B_U}$. As $\ket{\Psi}$ is
symmetric under exchange of the system and the ancilla, we have
\begin{align}
E[(U^{A_\anc B_\anc})^*\ket{\Psi}_A\ket{\Psi}_B] &=
E(U^*\ket{\Psi}_A\ket{\Psi}_B) \nn
& = E(U\ket{\Psi^*}_A\ket{\Psi^*}_B) \nn
& = E(U\ket{\Psi}_A\ket{\Psi}_B) \nn
& = E_U^{\Psi}.
\end{align}
In the second line we have used the fact that the entanglement is invariant
under complex conjugation, and in the third line we have used the fact that
$\ket{\Psi}_A=\ket{\Psi^*}_A$. Thus we obtain $E_U^{\Psi} = E_{U\dg}^{\Psi}$.

To show symmetry for $\Delta E_U^{\Psi}$, we simply use
\begin{align}
\Delta E_{U\dg}^{\Psi} &= \max_{U_0} [E_{U\dg U_0}^{\Psi} - E_{U_0}^{\Psi}] \nn
&= \max_{U_0} [E_{UU_0\dg}^{\Psi} - E_{U_0\dg}^{\Psi}] \nn
&= \max_{U_0} [E_{UU_0}^{\Psi} - E_{U_0}^{\Psi}] \nn
& = \Delta E_U^{\Psi}.
\end{align}
In the second line we have used $E_U^{\Psi} = E_{U\dg}^{\Psi}$. Thus we see that
these entangling capacities are symmetric, in contrast to $E_U$ and $\Delta E_U$
\cite{linden,aram}.

\subsection{Lower limit on Holevo capacities}
\label{sec:3b}
To relate the Holevo capacities to the entangling capacities, we first consider
an initial state $\ket{\Psi}_A\ket{\chi}_B$ that achieves the supremum in
Eq.\ \eqref{uni1}. From the reasoning in Ref.\ \cite{nielsen}, the ancilla for
Bob need have dimension no larger than $d_{B_U}$, so there exists a state that
achieves the supremum. In the following we take $d_{B_{\rm anc}}=d_{B_U}$.

We now add the additional subsystem for Bob $\h_{B_2}$ with dimension $d_{A_U}$,
and consider the state $\ket{\Psi}_{A_UB_2}\ket{\chi}_B$. That is, the state is
maximally entangled between $\h_{A_U}$ and $\h_{B_2}$, but the state is still
$\ket{\chi}_B$ on the subsystems $\h_{B_U}$ and $\h_{B_{\rm anc}}$. Now let
$V_j^{A_U}$ be the local unitary operators on $\h_{A_U}$ that transform
$\ket{\Psi}_{A_UB_2}$ to the $d_{A_U}^2$ mutually orthogonal maximally entangled
states. These operators satisfy $\sum_j V_j^{A_U} \rho {V_j^{A_U}}\dg \propto
\openone$. We take the initial ensemble to be
\begin{equation}
\e = \{ 1/d_{A_U}^2 , V_j^{A_U} \ket{\Psi}_{A_UB_2}\ket{\chi}_B \}.
\end{equation}

Applying the local operations $V_j^{A_U}$ is equivalent to applying
$(V_j^{B_2})^T$, where this denotes the transpose of $V_j^{A_U}$ acting on
$\h_{B_2}$. As local unitaries on $\h_{B_2}$ commute with $U$,
the states in the final ensemble are
\begin{equation}
(V_j^{B_2})^T U \ket{\Psi}_{A_UB_2}\ket{\chi}_B.
\end{equation}
The entropy of the
reduced density operator for $\h_{B}$ for the average state is then
\begin{align}
&S\left[\frac 1{d_{A_U}^2} \sum_{j=1}^{d_{A_U}^2} \Tr_{A_UB_2}
\left( U V_j^{A_U} \ket{\Psi}_{A_UB_2} \ket{\chi}_B\right) \right] \nn
&= S\left\{\frac 1{d_{A_U}^2}\sum_{j=1}^{d_{A_U}^2}\Tr_{A_UB_2}
[(V_j^{B_2})^T U \ket{\Psi}_{A_UB_2} \ket{\chi}_B]\right\} \nn
&= S[\Tr_{A_UB_2} (U \ket{\Psi}_{A_UB_2}\ket{\chi}_B)] =
E_U^{\Psi,\rightarrow}.
\end{align}

The action of the operators $V_j^{A_U}$ is to disentangle $\h_{A_U}$
and $\h_{B_2}$:
\begin{equation}
\frac 1{d_{A_U}^2}\sum_{j=1}^{d_{A_U}^2} V_j^{A_U} \ket{\Psi}_{A_UB_2}\bra{\Psi}
 (V_j^{A_U})\dg = \frac 1{d_{A_U}^2} \openone_{A_U}\otimes\openone_{B_2}.
\end{equation}
The operation $U$ does not act upon $\h_{B_2}$, so the average output state is
still a tensor product between $\h_{A_U}\otimes\h_B$ and $\h_{B_2}$. Therefore
the entropy of the complete reduced density operator for Bob is the sum of that
for $\h_{B}$, which is $E_U^{\Psi,\rightarrow}$, and that for $\h_{B_2}$, which
is $\log d_{A_U}$. That is,
\begin{align}
&S\left[\frac 1{d_{A_U}^2}\sum_{j=1}^{d_{A_U}^2} \Tr_{A_U} \left(U V_j^{A_U}
\ket{\Psi}_{A_UB_2}\ket{\chi}_B \right)\right] \nn
&= \log d_{A_U} + E_U^{\Psi,\rightarrow}.
\end{align}
Because Alice's subsystem has dimension $d_{A_U}$, the entanglement of the
individual pure states can not be more than $\log d_{A_U}$. Thus the total
Holevo information for the final ensemble is at least $E_U^{\Psi,\rightarrow}$,
so
\begin{equation}
E_U^{\Psi,\rightarrow} \le \chi_U^{\rightarrow}.
\end{equation}
Note that it is also possible for the entanglement of the individual pure
states to be less than $\log d_{A_U}$, so this reasoning shows that the Holevo
information for the ensemble is $\ge E_U^{\Psi,\rightarrow}$, rather than
exactly equal to $E_U^{\Psi,\rightarrow}$.

By applying exactly the same reasoning with the roles of Alice and Bob reversed,
we also have
\begin{equation}
E_U^{\Psi,\leftarrow} \le \chi_U^{\leftarrow}.
\end{equation}
In addition, since $E_U^{\Psi}\le E_U^{\Psi,*}$, we have
\begin{equation}
\label{Echi}
E_U^{\Psi} \le \chi_U^{*},
\end{equation}
where $*$ indicates either $\leftarrow$ or $\rightarrow$.

We can use similar reasoning to relate $\Delta E_U^{\Psi}$ and
$\Delta\chi_U^{*}$. Let us consider a $U_0$ that achieves the maximum in
Eq.\ \eqref{eq:difen}. Now we construct the initial ensemble
\begin{equation}
\e = \{ 1/d_{A_U}^2 , U_0 V_j^{A_U} \ket{\Psi}_{A_UB_2}\ket{\Psi}_B \},
\end{equation}
where $V_j^{A_U}$ acts upon $\h_{A_U}$. We may use the same approach as above to
determine the Holevo information of Bob's subsystem
for this ensemble. However, in this case the
entanglement of the individual pure states in the ensemble is exactly
$\log d_{A_U}$. Therefore, this ensemble has Holevo information of exactly
$E_{U_0}^{\Psi}$. Applying $U$ to this
ensemble then gives an ensemble with Holevo information $E_{UU_0}^{\Psi}$.
The change in Holevo information is therefore $E_{UU_0}^{\Psi}-E_{U_0}^{\Psi}$.
When we take the $U_0$ that gives the maximum in Eq.\ \eqref{eq:difen},
the change in Holevo information is exactly equal to $\Delta E_U^{\Psi}$. Hence
we have
\begin{equation}
\Delta E_U^{\Psi} \le \Delta\chi_U^{\rightarrow}.
\end{equation}
Using the same reasoning with the roles of Alice and Bob reversed yields
$\Delta E_U^{\Psi} \le \Delta\chi_U^{\leftarrow}$.

This reasoning is not sufficient to show $\Delta E_U^{\Psi,*}\le
\Delta\chi_U^{*}$, because the initial ensemble can have Holevo information
larger than $E_{U_0}^{\Psi,*}$. Note that if it were possible to show
$\Delta E_U^{\Psi,*}\le\Delta\chi_U^{*}$, that would imply
$\Delta E_U\le\Delta\chi_U^{*}$. This is because, if one allows larger
dimensions of $\h_{A_U}$ and $\h_{B_U}$ for a given $U$, $\Delta\chi_U^{*}$
is unchanged, but $\Delta E_U^{\Psi,*}=\Delta E_U$ (as discussed in
Sec.\ \ref{sec:ent}). A counterexample to $\Delta E_U\le\Delta\chi_U^{*}$ was
given by Ref.\ \cite{aram}, so it is unsurprising that this line of reasoning
does not allow us to prove $\Delta E_U^{\Psi,*}\le\Delta\chi_U^{*}$.
Note that, because we have proven $\Delta E_U^{\Psi}\le\Delta\chi_U^{*}$, the
capacity $\Delta E_U^{\Psi}$ can not simplify to $\Delta E_U$ under an expanded
dimension.

\section{Channels}
\label{sec:cha}
We can alternatively derive some of the above results using results for quantum
channels. For a given state $\ket{\chi}$, we may define a quantum channel as
\begin{equation}
\Phi_{\ket{\chi}}(\rho_A) \equiv
 \Tr_A [U(\rho_A \otimes \ket{\chi}_B \bra{\chi})U\dg ].
\end{equation}
The capacity of this channel without entanglement assistance is given by
\begin{equation}
C(\Phi_{\ket{\chi}}) = \sup_\e \chi [\Phi_{\ket{\chi}}(\e)].
\end{equation}
The capacity $\chi_U'^{\rightarrow}$ is then given by
\begin{equation}
\chi_U'^{\rightarrow} = \sup_{\ket{\chi}} C(\Phi_{\ket{\chi}}).
\end{equation}
Thus we can see that there is a fundamental connection between these two
capacities. There is a similar connection for communication from Bob to Alice.

We may also show a connection between the Holevo capacities of $U$ and
the entanglement assisted capacity of a quantum channel. Let us define a
Holevo capacity by
\begin{equation}
\chi_U^{C,\rightarrow} \equiv \sup_{ p_j, V_j^{A}, \ket{\phi}_{AB_2},
\ket{\chi}_{B} }\chi ( p_j, \Tr_A U V_j^{A} \ket{\phi}_{AB_2}\ket{\chi}_{B} ),
\end{equation}
where $V_j^{A}$ is a unitary acting on $\h_{A}$.
Here $\h_{B_2}$ is an additional subsystem for Bob. Initially Alice's subsystem
may be entangled with $\h_{B_2}$, but not with $\h_{B}$. This capacity is
related to the entanglement assisted capacity of the channel
$\Phi_{\ket{\chi}}$. In particular,
\begin{equation}
C_E(\Phi_{\ket{\chi}}) = \sup_{ p_j, V_j^{A}, \ket{\phi}_{AB_2}}
\chi ( p_j, \Tr_A U V_j^{A} \ket{\phi}_{AB_2}\ket{\chi}_{B} ),
\end{equation}
so
\begin{equation}
\label{rel}
\chi_U^{C,\rightarrow} = \sup_{\ket{\chi} }C_E(\Phi_{\ket{\chi}}).
\end{equation}
It is clear from the definition of the capacity $\chi_U^{C,\rightarrow}$ that it
is less restrictive than $\chi_U'^{\rightarrow}$, but it is more restrictive
than $\chi_U^{\rightarrow}$. Thus $\chi_U'^{\rightarrow}\le
\chi_U^{C,\rightarrow}\le\chi_U^{\rightarrow}$.

We can also obtain the result \eqref{Echi} using the standard formula for the
entanglement assisted channel capacity. Recall that the channel capacity is
given by \cite{bsst}
\begin{equation}
\label{formula}
C_E(\Phi_{\ket{\chi}}) = \max_{\rho\in\h_{A_U}} S(\rho)+
S(\Phi_{\ket{\chi}}(\rho))-S(\Phi_{\ket{\chi}}\otimes
\openone(\ket{\psi}\bra{\psi})),
\end{equation}
where $\ket{\psi}$ is a pure state with $\rho$ as the reduced density operator.
Now take $\ket{\chi}$ to be the state that achieves the supremum in
Eq.\ \eqref{uni1}. Let $\rho$ be the maximally mixed state in $\h_{A_U}$, and
let $\ket{\psi}$ be a maximally entangled state between $\h_{A_U}$ and
$\h_{A_{\rm anc}}$. Then
\begin{align}
S(\rho) &= \log d_{A_U} \\
S(\Phi_{\ket{\chi}}(\rho)) &= E_U^{\Psi,\rightarrow}.
\end{align}
The entropy $S(\Phi_{\ket{\chi}}\otimes I(\ket{\psi}\bra{\psi}))$ is the entropy
of the subsystem $\h_{A_U}$ after applying $U$, and can not exceed
$\log d_{A_U}$. Thus we find that, for this choice of $\ket{\chi}$,
$C_E(\Phi_{\ket{\chi}})\ge E_U^{\Psi,\rightarrow}$. Because we also have
Eq.\ \eqref{rel}, this implies that $E_U^{\Psi,\rightarrow}\le
\chi_U^{\rightarrow}$. This method of proof is closely related to the more
explicit scheme given in the preceding section. The proof of the capacity
formula \eqref{formula} yields an ensemble equivalent to what we have given
above. We may also use the same reasoning for communication from Bob to Alice,
to obtain the inequality $E_U^{\Psi,\leftarrow}\le\chi_U^{\leftarrow}$

\section{Generali{\s}ations}
\label{sec:imp}
With these results, we have the set of inequalities between the entangling and
Holevo capacities summari{\s}ed in Table \ref{table}. The inequality
$\Delta E_U\le\Delta\chi_U^*$ was shown in Refs.\ \cite{berry1,berry2}, and the
inequality $E_U \le \chi_U^*$ was shown conditionally upon numerical results in
Ref.\ \cite{berry3}. The two-qudit inequalities that we have proven here are
weaker than those for the two-qubit case, and it is natural to ask if the
two-qubit inequalities may be extended to the two-qudit case.
Ref.\ \cite{berry3} showed that the inequalities may be extended to the
two-qudit case for controlled unitaries. Here we present a different approach
to generali{\s}ing these inequalities.

\begin{table*}
\caption{The inequalities that have been shown for two-qubit and two-qudit
systems.\label{table}} \begin{tabular}{|c|c|c|} \hline
2-qubit                       & 2-qudit                            \\ \hline
$E_U \le \chi_U^*$            & $E_U^{\Psi,*} \le \chi_U^*$        \\
$\Delta E_U\le\Delta\chi_U^*$ & $\Delta E_U^\Psi\le\Delta\chi_U^*$ \\
\hline
\end{tabular}
\end{table*}

\subsection{Alternative ensembles}
The ensembles given in the previous sections do not generali{\s}e to initial states
that are not maximally entangled. This is because we rely on a unitary operation
on one part of the entangled state being equivalent to a unitary on the other
half of the entangled state. We therefore present an alternative method of
producing ensembles that is more complicated, but is more easily generali{\s}ed.

As in Sec.\ \ref{sec:3b} we use the state $\ket{\Psi}_A\ket{\chi}_B$ that
achieves the supremum in Eq.\ \eqref{uni1}. We take $V_j^{A_\anc}$ to be
local unitaries acting upon $\h_{A_\anc}$ that transform $\ket\Psi_A$ to the
$d_{A_U}^2$ mutually orthogonal maximally entangled states.
Now consider the ensemble
\begin{equation}
\e = \left\{ 1/d_{A_U}^2 , \frac 1{d_{A_U}} \sum_{j=1}^{d_{A_U}^2}
e^{i 2\pi kj/d_{A_U}^2} V_j^{A_\anc} \ket{\Psi}_A \ket{\chi}_B
\ket{j}_{B_2} \right\}.
\end{equation}
Because the states $V_j^{A_\anc}\ket\Psi_A$ are orthogonal, the states in the
ensemble may be obtained by local unitaries on $\h_A$.

Using the notation $\ket\omega_{AB}=U\ket{\Psi}_A\ket{\chi}_B$, the entropy
of the average state for $\h_B\otimes\h_{B_2}$ after applying $U$ is
\begin{align}
&S\left[\frac 1{d_{A_U}^2}\Tr_A \left( \sum_{j=1}^{d_{A_U}^2} V_j^{A_\anc}
\ket\omega_{AB}\bra\omega (V_j^{A_\anc})\dg \otimes \ket j_{B_2}\bra j
\right)\right] \nn &= S\left[\Tr_A  \ket\omega_{AB}\bra\omega\otimes
\left(\frac 1{d_{A_U}^2}\openone_{B_2} \right)\right] \nn
&= E_U^{\Psi,\rightarrow} + 2\log d_{A_U}.
\end{align}
Because the entropy of the individual states can not exceed $2\log d_{A_U}$,
the Holevo information must be at least $E_U^{\Psi,\rightarrow}$. This again
gives the inequality $E_U^{\Psi,*} \le \chi_U^*$.

As in Sec.\ \ref{sec:3b}, we can not ensure that the entropy of the individual
states is exactly $2\log d_{A_U}$, so we can not show that
$\Delta E_U^{\Psi,*} \le \Delta \chi_U^*$. However, it is straightforward to
show the other inequalities shown in Sec.\ \ref{sec:3b} using this
alternative approach.

The advantage of this approach is that we may use it for states that are not
maximally entangled with the ancilla. Let $\ket{\phi}_A\ket{\chi}_B$ be the
initial state such that the maximal entanglement $E_U$ is obtained under $U$.
Now let $\{V_j^A\}$ be a set of $d_{A_U}^2$ unitaries such that
\begin{equation}
_A\bra\phi {V_j^A}\dg V_k^A\ket\phi_A = \delta_{jk}.
\end{equation} 
It is easily shown that (taking $d_{A_\anc}=d_{A_U}$) this condition is equivalent to
\begin{equation}
\label{con1}
\sum_{j=1}^{d_{A_U}^2} V_j^A\ket\phi_A\bra\phi {V_j^A}\dg = \openone_A.
\end{equation} 
We also require unitaries $V_j^B$, $U_j^A$, and $U_j^B$ such that
\begin{equation}
\label{con2}
UV_j^A\otimes V_j^B = U_j^A\otimes U_j^B U.
\end{equation} 
If there exists such a set of unitaries, we may construct the ensemble
\begin{equation}
\e = \left\{ 1/d_{A_U}^2 , \frac 1{d_{A_U}} \sum_{j=1}^{d_{A_U}^2}
e^{i 2\pi kj/d_{A_U}^2} V_j^A \ket{\phi}_A V_j^B \ket{\chi}_B \ket{j}_{B_2} \right\}.
\end{equation}

As before, because the states $V_j^A \ket{\phi}_A$ are orthogonal, the states
in the ensemble may be obtained by local unitaries in $\h_A$. The entropy
of the average state for $\h_B\otimes\h_{B_2}$ after applying $U$ is then
\begin{align}
&S\left[\frac 1{d_{A_U}^2}\Tr_A \left( \sum_{j=1}^{d_{A_U}^2} U_j^A\otimes U_j^B
\ket\omega_{AB}\bra\omega (U_j^A\otimes U_j^B)\dg \right.\right. \nn &
\left.\left. \phantom{\sum_{j=1}^{d_{A_U}^2}} \otimes \ket j_{B_2}\bra j \right)\right] \nn
&= S\left[\frac 1{d_{A_U}^2} \sum_{j=1}^{d_{A_U}^2} U_j^B \Tr_A (\ket\omega_{AB}\bra\omega)
{U_j^B}\dg \otimes \ket j \bra j \right] \nn
&= E_U^{\Psi,\rightarrow} + 2\log d_{A_U},
\end{align}
where $\ket\omega = U\ket{\phi}_A\ket{\chi}_B$.
Thus we find that, provided there exists a set of unitaries satisfying
conditions \eqref{con1} and \eqref{con2},
\begin{equation}
\label{ineq1}
E_U \le \chi_U^\rightarrow.
\end{equation}
It is trivially seen that there is an equivalent set of conditions for 
$E_U \le \chi_U^\leftarrow$, simply by reversing the roles of Alice and Bob.

There are a number of cases where conditions \eqref{con1} and \eqref{con2}
may be satisfied. The first is where $\ket\phi$ is a maximally entangled state,
as above. Then we simply take $V_j^A=U_j^A$ and $V_j^B=U_j^B=\openone$. Another
case is for two-qubit unitaries if
\begin{equation}
\label{eq:phicon}
\ket{\phi}_A = \lambda_0 \ket{\psi_0}_{A_\anc}\ket{0}_{A_U} +
\lambda_1 \ket{\psi_1}_{A_\anc}\ket{1}_{A_U},
\end{equation}
where the $\lambda_i$ are real, $\lambda_0^2+\lambda_1^2=1$, and
$\braket{\psi_0}{\psi_1}$ is real. It is not possible to give an arbitrary
two-qubit state $\ket{\phi}_A$ in this form, as the inner product may have an
imaginary component.
We assume that the two-qubit unitary has been simplified to the canonical form
\cite{kraus}
\begin{equation}
\label{canon}
U_d = \exp \left(-i\sum_{j=1}^3 \alpha_j \sigma_j\otimes \sigma_j \right).
\end{equation}

In this case we define local unitaries $V_a$ and $V_b$ such that
\begin{align}
V_a\ket{\psi_0}_{A_\anc} &= \ket{\psi_1}_{A_\anc}, \nn
V_a\ket{\psi_1}_{A_\anc} &= \ket{\psi_0}_{A_\anc}, \nn
V_b\ket{\psi_0}_{A_\anc} &= \ket{\psi_0^\perp}_{A_\anc}, \nn
V_b\ket{\psi_1}_{A_\anc} &= \ket{\psi_1^\perp}_{A_\anc},
\end{align}
where the superscript $\perp$ indicates perpendicular states. We then take
\begin{align}
V_{jk}^A &= (V_b)^k (V_a)^j \otimes \sigma_y^j, \nn
V_{jk}^B &= \sigma_y^j \otimes \openone,
\end{align}
for $j,k\in\{0,1\}$, and $V_{jk}^A=U_{jk}^A$ and $V_{jk}^B=U_{jk}^B$. It is
easily seen that the conditions \eqref{con1} and \eqref{con2} are
satisfied (where we have replaced the single index ``$j$'' with ``$j,k$'').

Thus we find that, provided it is possible to give $\ket{\phi}_A$ in the form
\eqref{eq:phicon} in the two-qubit case, then $E_U \le \chi_U^\rightarrow$.
In the case that $\braket{\psi_0}{\psi_1}$ is not real, then
we have the problem that $V_a$ is not unitary, so the derivation does not apply.
Numerically it is found that it is always possible to give $\ket{\phi}_A$ in the
form \eqref{eq:phicon}. This gives an alternative proof of $E_U \le
\chi_U^\rightarrow$ to that in Ref.\ \cite{berry3}, though it still depends on
numerical results.

\subsection{Generali{\s}ation of $\Delta E_{U}\le\Delta\chi_U^\rightarrow$}
We may also generali{\s}e the proof of $\Delta E_U\le\Delta\chi_U^*$ for two-qubits
from \cite{berry1,berry2}. Let $V_j^{A_U}$ be a set of $d_{B_U}^2$ unitary operators
on $\h_{A_U}$ such that
\begin{equation}
\label{con3}
\sum_{j=1}^{d_{A_U}^2} {V_j^{A_U}}\dg \rho V_j^{A_U} \propto \openone_{A_U}.
\end{equation}
We also require unitaries $V_j^{B_U}$, $U_j^{A_U}$ and $U_j^{B_U}$ satisfying
\begin{equation}
\label{con4}
U V_j^{A_U}\otimes V_j^{B_U} = U_j^{A_U}\otimes U_j^{B_U} U.
\end{equation}

For any $\epsilon>0$, we may select an initial state $\ket\psi$ such that $U\dg$
decreases the entanglement by $\Delta E_U-\epsilon$. For the ensemble
\begin{equation}
\e=\{1/d_{A_U}^2,{U_j^{A_U}}\dg \otimes {U_j^{B_U}}\dg \ket\psi\},
\end{equation}
the increase in the Holevo information under $U$ is
\begin{align}
& S\left[ \frac 1{d_{A_U}^2} \sum_j {V_j^{A_U}}\dg \Tr_B (U\dg\ket\psi)V_j^{A_U}\right]
-S\left[ \Tr (U\dg\ket\psi)\right] \nn
&\quad -\left\{ S\left[ \frac 1{d_{A_U}^2} \sum_j {U_j^{A_U}}\dg \Tr_B (\ket\psi)U_j^{A_U}\right]
-S\left[ \Tr (\ket\psi)\right]\right\} \nn
&= S\left[ \Tr_{A_UB} (U\dg\ket\psi)\otimes (\openone_{A_U}/d_{A_U}) \right] \nn
&\quad -S\left[ \frac 1{d_{A_U}^2} \sum_j {U_j^{A_U}}\dg \Tr_B (\ket\psi)U_j^{A_U}\right]
+\Delta E_U-\epsilon \nn
& \ge \Delta E_U-\epsilon.
\end{align}
Hence, provided we have a set of unitaries satisfying the restrictions
\eqref{con3} and \eqref{con4}, $\Delta E_U-\epsilon\le\Delta
\chi_{U\dg}^\leftarrow$. As this is true for all $\epsilon>0$, we obtain
$\Delta E_U\le\Delta\chi_{U\dg}^\leftarrow$. Because
$\Delta\chi_U^\leftarrow=C_\leftarrow^E(U)$ \cite{bhls} and
$C_\leftarrow^E(U)=C_\rightarrow^E(U\dg)$ \cite{aram},
$\Delta E_{U}\le\Delta\chi_{U}^\rightarrow$.

Thus we find that, provided there exists a set of unitaries $V_j^{A_U}$,
$V_j^{B_U}$, $U_j^{A_U}$ and $U_j^{B_U}$ satisfying Eqs.\ \eqref{con3} and
\eqref{con4}, then
\begin{equation}
\label{ineq2}
\Delta E_{U}\le\Delta\chi_U^\rightarrow.
\end{equation}
This proof is a generali{\s}ation of that for two qubits in
Refs.\ \cite{berry1,berry2}, where $V_j^{A_U}=V_j^{B_U}=U_j^{A_U}=U_j^{B_U}
=\sigma_j$.

The conditions required for the proof of $\Delta E_{U}\le\Delta\chi_U^\rightarrow$
to hold, \eqref{con3} and \eqref{con4}, are remarkably similar to those required
for the proof of $E_{U}\le\chi_U^\rightarrow$, \eqref{con1} and \eqref{con2}.
The main difference is that the operators in the conditions \eqref{con1} and
\eqref{con2} are on the entire space $\h_A$, which is of dimension $d_{A_U}^2$.
Because there are only $d_{A_U}^2$ operators $V_j^A$, \eqref{con1} only applies
to one state $\ket\phi$, rather than all states.

\subsection{Implications of Refs.\ \cite{aram,linden}}
Although there are cases of two-qudit unitaries $U$ for which one can find
unitaries satisfying \eqref{con1} and \eqref{con2} or \eqref{con3}
and \eqref{con4}, it can not be possible to find such unitaries for all $U$.
This follows from the results of Ref.\ \cite{aram} which shows that there are
examples of unitaries that violate the inequalities \eqref{ineq1} and \eqref{ineq2}.

Ref.\ \cite{aram} gives the example of a series of unitaries, $V_m$, such that
$C_\leftarrow^E(V_m)\le O(\log^2 m)$, but $\Delta E_{V_m}=m$. Because
$C_\leftarrow^E(V_m)=\Delta\chi_{V_m}^\leftarrow$ \cite{bhls}, this unitary
gives an example of a violation of the inequality \eqref{ineq2} (with subsystems
$\h_A$ and $\h_B$ reversed). These unitaries also satisfy $E_{V_m}=m$; because
$\chi_{V_m}^\leftarrow \le \Delta\chi_{V_m}^\leftarrow$, these unitaries also
satisfy $\chi_{V_m}^\leftarrow < E_{V_m}$. Thus, the inequality \eqref{ineq1}
can also be violated for two-qudit unitaries.

Part of the significance of the results in Ref.\ \cite{aram} is that there is
an exponential separation between the entanglement capacity and communication
(or Holevo) capacity. This demonstrates that the close numerical relationship
observed for two-qubit operations \cite{berry3} does not hold in higher
dimensions. However, for the new entanglement capacities that have been defined
here, exponential separations are not obtained in many cases. For $U$ equal to the operation
$V_m$ of Ref.\ \cite{aram}, it is easily seen that $E_U^{\Psi,\rightarrow}=
\chi_U^{\rightarrow}=m$. In addition, because
\begin{equation}
E_U^{\Psi,\leftarrow}\le\chi_U^{\leftarrow}\le \Delta\chi_U^{\leftarrow}=
C_\leftarrow^E(U) \le O(\log^2 m)
\end{equation}
there is no exponential separation between $E_U^{\Psi,\leftarrow}$ and
$\chi_U^{\leftarrow}$. Similarly, we have $\Delta E_U^{\Psi,\rightarrow}=
\Delta\chi_U^{\rightarrow}=m$.

We find exponential separation between $\Delta E_U^\Psi$ and
$\Delta\chi_U^\rightarrow$, and between $E_U^\Psi$ and $\chi_U^\rightarrow$. In
each case, the entangling capacity is polynomial in $\log m$, whereas the Holevo
capacity is $m$. We also have exponential separation between
$\Delta E_U^{\Psi,\leftarrow}$ and $\Delta\chi_U^{\leftarrow}$ provided we allow
expanded dimensions for $\h_{A_U}$ and $\h_{B_U}$ (so
$\Delta E_U^{\Psi,\leftarrow}=\Delta E_U$). If we do not allow expanded
dimensions, we don't have sufficient information to determine if there is
exponential separation.

\begin{figure}
\centering
\includegraphics[width=0.45\textwidth]{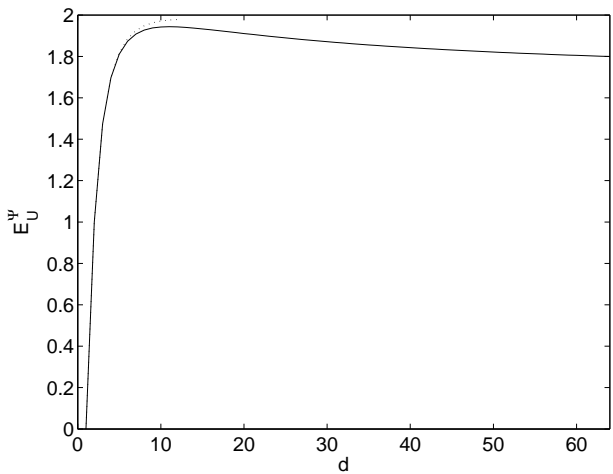}
\caption{The values of $E_U^{\Psi}$ (solid line) and $E_U^{\Psi,\leftarrow}$
(dotted line) for the unitary given in Ref.\ \cite{aram} and various subsystem
dimensions $d$ (in \cite{aram} $d=2^m$).}
\label{entang}
\end{figure}

Another interesting aspect of these results is that there is exponential
separation between $E_U^{\Psi}$ and $E_U$. In fact, numerical calculations
indicate that $E_U^{\Psi}$ does not exceed 2 for these states (see
Fig.\ \ref{entang}), whereas $E_U=m$. Recall that the state
$U\ket{\Psi}_A\ket{\Psi}_B$ in the definition of $E_U^{\Psi}$ may be used to
implement the unitary $U$ given the correct results for local Bell measurements.
If this method for implementing $U$ is applied to the correct initial state,
then entanglement of $m$ will be obtained, despite the entangled resource state
having low entanglement.

We may also use the results of Ref.\ \cite{linden} to deduce results for
bipartite unitaries with maximum entanglement capability. From
Ref.\ \cite{linden}, if a unitary on subsystems
of dimensions $d_{A_U}$ and $d_{B_U}$ produces entanglement of $2\log d_{A_U}$
(for $d_{A_U}\le d_{B_U}$), then this entanglement may be produced for an
initial state of the form $\ket{\Psi}_A\ket{\chi}_B$. Thus
$E_U^{\Psi,\rightarrow}=\Delta E_U=2\log d_{A_U}$. From the results given above,
we also have $\chi_U^{\rightarrow}=2\log d_{A_U}$. Similarly, for
$d_{A_U}\ge d_{B_U}$, if $\Delta E_U=2\log d_{B_U}$ then
$\chi_U^{\leftarrow}=2\log d_{B_U}$.

\section{Conclusions}
\label{sec:con}
In the case of two-qubit unitaries, it is known that the inequalities between
entanglement capacity and Holevo capacity $\Delta E_U\le\Delta\chi_U^*$ and
$E_U \le \chi_U^*$ hold \cite{berry1,berry2,berry3}. The methods used to prove
these inequalities use results that are specific to the case of qubits, and can
not be generali{\s}ed to the case of qudits (except for restricted classes of
unitaries). In addition, Ref.\ \cite{aram} shows
that these inequalities can be violated for two-qudit unitaries.

Here we have shown that if alternative definitions of the entangling
capacities based on the Jamio{\l}kowski isomorphism \cite{jam} are used, then
we can obtain analogous inequalities for all two-qudit unitaries. In addition,
we have shown that there is a close connection between these inequalities and
the theory of entanglement assisted channel capacities.

These results complement those of Ref.\ \cite{berry3}, where inequalities
between a restricted form of the Holevo capacity and the entangling capacity
were proven. Together with the results shown here, we obtain the set of
inequalities
\begin{equation}
\begin{array}{*{20}c}
E_U^\Psi\le \chi_U^* \\
\begin{sideways}{$\ge$}
\end{sideways} \hspace{6 mm}
\begin{sideways}{$\le$}
\end{sideways} \\
E_U\ge \chi_U'^{*}
\end{array}
\hspace{2 cm}
\begin{array}{*{20}c}
\Delta E_U^\Psi\le \Delta\chi_U^* \\
\begin{sideways}{$\ge$}
\end{sideways} \hspace{8 mm}
\begin{sideways}{$\le$}
\end{sideways} \\
\Delta E_U\ge \Delta\chi_U'^{*}
\end{array}
\end{equation}
These inequalities do not give an ordering between $\Delta\chi_U^*$ and
$\Delta E_U$ or between $\chi_U^*$ and $E_U$. This is not possible,
because there are examples of cases where $\Delta E_U<\Delta\chi_U^*$
and $E_U<\chi_U^*$ \cite{berry3}, and examples where
$\Delta E_U>\Delta\chi_U^*$ and $E_U>\chi_U^*$ \cite{aram}.

These results demonstrate that we can not generali{\s}e the method of proof used
here to show the corresponding inequalities for $E_U$ and $\Delta E_U$.
However, we have given conditions on the unitary such that the inequalities
$E_{U}\le\chi_U^\rightarrow$ or $\Delta E_{U}\le\Delta\chi_U^\rightarrow$ are
satisfied. Although the conditions in the two cases are not identical, they are
very similar. This approach gives a more elegant method of showing
$E_{U}\le\chi_U^\rightarrow$ in the two-qubit case than that used in
Ref.\ \cite{berry3}.

Although exponential separations between entanglement and communication
capacities were demonstrated in Ref.\ \cite{aram}, for the example in
Ref.\ \cite{aram} there are not exponential separations between
$E_U^{\Psi,*}$ and $\chi_U^*$. Whether there are cases of unitaries for
which there is exponential separation between these capacities is an interesting
topic for future study.

\acknowledgments
The author acknowledges valuable discussions with Jason Twamley, David Bulger
and Barry Sanders. This project has been supported by the Australian Research
Council.


\begin{thebibliography}{}
\bibitem{cirac} J. I. Cirac, W. D\"ur, B. Kraus, and M. Lewenstein, \prl
\textbf{86}, 544 (2001).
\bibitem{dur} W. D\"ur and J. I. Cirac, \pra \textbf{64}, 012317 (2001).
\bibitem{implement} D. W. Berry, \pra \textbf{75}, 032349 (2007).
\bibitem{eisert} J. Eisert, K. Jacobs, P. Papadopoulos, and M. B. Plenio,
\pra \textbf{62}, 052317 (2000).
\bibitem{collins} D. Collins, N. Linden, and S. Popescu, \pra \textbf{64},
032302 (2001).
\bibitem{zanardi} P. Zanardi, C. Zalka, and L. Faoro, \pra \textbf{62},
030301(R) (2000).
\bibitem{durvid} W. D\"ur, G. Vidal, J. I. Cirac, N. Linden, and S. Popescu,
\prl \textbf{87}, 137901 (2001).
\bibitem{kradur} B. Kraus, W. D\"ur, G. Vidal, J. I. Cirac, M. Lewenstein,
N. Linden, and S. Popescu, Z. Naturforsch \textbf{56a}, 91 (2001).
\bibitem{kraus} B. Kraus and J. I. Cirac, \pra \textbf{63}, 062309 (2001).
\bibitem{leifer} M. S. Leifer, L. Henderson, and N. Linden,
\pra \textbf{67}, 012306 (2003).
\bibitem{childs} A. M. Childs, D. W. Leung, F. Verstraete, and G. Vidal,
Quantum Information and Computation \textbf{3}, 97 (2003).
\bibitem{kraus2} B. Kraus, K. Hammerer, G. Giedke and J. I. Cirac,
\pra \textbf{67}, 042314 (2003).
\bibitem{wang} X. Wang, B. C. Sanders, and D. W. Berry, \pra \textbf{67}, 042323 (2003).
\bibitem{bravyi} S. Bravyi, arXiv:0704.0964v1 [quant-ph].
\bibitem{groisman} B. Groisman, arXiv:0704.1042v1 [quant-ph].
\bibitem{bhls} C. H. Bennett, A. W. Harrow, D. W. Leung, and J. A. Smolin,
IEEE Trans. Inf. Theory \textbf{49}, 1895 (2003).
\bibitem{berry1} D. W. Berry and B. C. Sanders, \pra \textbf{67}, 040302(R)
(2003).  
\bibitem{berry2} D. W. Berry and B. C. Sanders, \pra \textbf{68}, 032312 (2003).
\bibitem{berry3} D. W. Berry and B. C. Sanders, \pra \textbf{71}, 022304 (2005). 
\bibitem{linden} N. Linden, J. A. Smolin, and A. Winter, quant-ph/0511217
(2005).
\bibitem{aram} A. W. Harrow and P. W. Shor, quant-ph/0511219 (2005).
\bibitem{jam} A. Jamio{\l}kowski, Rep. Math. Phys. \textbf{3}, 275 (1972).
\bibitem{kholevo} A. S. Kholevo, Probl. Peredachi Inf. \textbf{9}, 177 (1973).
\bibitem{holevo} A. S. Holevo, IEEE Trans. Inf. Theory \textbf{44}, 269 (1998).
\bibitem{shuwes} B. Schumacher and M. D. Westmoreland, \pra \textbf{56},
131 (1997).
\bibitem{nielsen} M. A. Nielsen, C. M. Dawson, J. L. Dodd, A. Gilchrist,
D. Mortimer, T. J. Osborne, M. J. Bremner, A. W. Harrow, and A. Hines, \pra
\textbf{67} 052301 (2003).
\bibitem{bsst} C. H. Bennett, P. W. Shor, J. A. Smolin, and A. V. Thapliyal, 
\prl \textbf{83}, 3081 (1999).
\end{thebibliography}
\end{document}